# On Social-Temporal Group Query with Acquaintance Constraint

De-Nian Yang
Academia Sinica
dnyang@iis.sinica.edu.tw

Yi-Ling Chen
National Taiwan Univ.
yiling@arbor.ee.ntu.edu.tw

Wang-Chien Lee
The Penn State Univ.
wlee@cse.psu.edu

Ming-Syan Chen
National Taiwan Univ.
mschen@cc.ee.ntu.edu.tw

## ABSTRACT

Three essential criteria are important for activity planning, including: (1) finding a group of attendees familiar with the initiator, (2) ensuring each attendee in the group to have tight social relations with most of the members in the group, and (3) selecting an activity period available for all attendees. Therefore, this paper proposes *Social-Temporal Group Query* to find the activity time and attendees with the minimum total social distance to the initiator. Moreover, this query incorporates an acquaintance constraint to avoid finding a group with mutually unfamiliar attendees. Efficient processing of the social-temporal group query is very challenging. We show that the problem is NP-hard via a proof and formulate the problem with Integer Programming. We then propose two efficient algorithms, *SGSelect* and *STGSelect*, which include effective pruning techniques and employ the idea of pivot time slots to substantially reduce the running time, for finding the optimal solutions. Experimental results indicate that the proposed algorithms are much more efficient and scalable. In the comparison of solution quality, we show that STGSelect outperforms the algorithm that represents manual coordination by the initiator.

## 1. INTRODUCTION

Three essential criteria are important for activity initiation, including: (1) finding a group of attendees familiar with the initiator, (2) ensuring each attendee in the group to have tight social relations with most of the members in the group, and (3) selecting an activity period available for all attendees. For example, if a person has a given number of complimentary tickets for a movie and would like to invite some friends, she usually prefers choosing a set of mutually close friends and the time that all of them are available. Nowadays, most activities are still initiated manually via phone, e-mail, messenger, etc. However, a growing number of systems are able to gather and make available some information required for activity initiation. For example, social networking websites, such as Facebook [2] and LinkedIn [5], provide the social relations and contact information, and web collaboration tools, such as Google Calendar [3], allow people to share their available time to friends and co-workers. Even with the availability of the above information, nevertheless, an activity initiator still needs to manually decide the suitable time and desirable attendees for activity planning, which can be very tedious and time-consuming. Thus, it is desirable to provide an efficient *activity planning service* that automatically suggests the attendees and time for an activity.

To support the aforementioned service, we formulate a new query problem, named *Social-Temporal Group Query (STGQ)*, which considers the available time of candidate attendees and their social relationship. Given an activity initiator, we assume that friends on her social network (i.e., her friends together with friends of friends) are candidate attendees. We also assume that the available time slots of the candidate attendees have been made available for the system in charge of query processing (e.g., via web collaboration tools with the corresponding privacy setting[1]), and the social relationship between friends is quantitatively captured as *social distance* [10, 12, 13], which measures the closeness between friends. An STGQ comes with the following parameters, (1) the activity size, denoted by $p$, to specify the number of expected attendees, (2) the length of time for the activity, denoted by $m$, (3) a social radius constraint, denoted by a number $s$, to specify the scope of candidate attendees, and (4) an acquaintance constraint, denoted by a number $k$, to govern the social relationship between attendees. The primary goal is to find a set of activity attendees and a suitable time which match the specified number of attendees in (1) and the activity length in (2) respectively, such that the total social distance between the initiator and all invited attendees is minimized. Additionally, the social radius constraint in (3) specifies that all the candidate attendees are located no more than $s$ edges away from the initiator on her social network, while the acquaintance constraint in (4) requires that each attendee can have at most $k$ other unacquainted attendees.[2] As such, by controlling $s$ and $k$ based on desired social atmosphere, suitable attendees and time are returned. Example 1 in the Appendix A shows an illustrative example. Note that, in situations where the time of the planned activity is pre-determined (e.g., complimentary tickets for a concert or a sport event in a specified day), the aforementioned STGQ can be simplified as a *Social Group Query (SGQ)* (i.e., without considering the temporal constraint). In this paper, we first examine the processing

---

[1]To process STGQ, it is not necessary for a user to share the schedule to friends. However, we assume that any friend can initiate an STGQ, and the query processing system can look up the available time of the user, just like the friend making a call to ask the available time. Different privacy policies or different schedules can be set for different friends, just like answering different available time, or even not answering, for various friends. The above privacy and schedule setting is beyond the scope of this paper.

[2]Note that the social radius constraint is specified in terms of *number of edges*, rather than the social distance, so the users can easily specify the constraint.

strategies for SGQ and then take further steps to extend our study on the more complex STGQ.

For SGQ, a simple approach is to consider every possible $p$ attendees and find the total social distance. This approach needs to evaluate $C_{p-1}^{f-1}$ candidate groups, where $f$ is the number of candidate attendees. In the current social network [8], the average number of friends for a user is around 100. Therefore, if an initiator would like to invite 10 attendees out of her 100 friends, the number of candidate groups is in the order of $10^{13}$. Note that the above scenario considers only the friends of the user as candidate attendees, and the number of candidate groups will significantly increase if the friends of each friend are also included. To sequentially choose the attendees at each iteration, giving priority to close friends based on their social distances to the activity initiator may lead to a smaller total social distance in the beginning, but does not always end up with a solution that satisfies the acquaintance constraint, especially for an activity with a small $k$. On the other hand, giving priority to a set of people who know each other to address the acquaintance constraint does not guarantee to find the solution with the minimum total social distance. Therefore, to select each attendee, the challenge comes from the dilemma of (1) reducing the total social distance at each iteration and (2) ensuring that the solution eventually follows all constraints of this problem.

Based on the above observations, in this paper, we first prove that SGQ is NP-hard and then propose an algorithm, called *SGSelect*, to find the optimal solution. We design three strategies, access ordering, distance pruning, and acquaintance pruning, to effectively prune the search space and reduce the processing time. During the selection of each attendee, we address both the social distance and connectivity of the attendee, together with the characteristics of the social network for other attendees that have not been considered, in order to find the optimal solution. Moreover, to consider the temporal dimension, we formulate STGQ as an optimization problem and derive an Integer Programming optimization model, which can be modified to support SGQ as well. Processing STGQ is significantly more challenging than processing SGQ because there may exist numerous time periods with different candidate groups. To solve the problem, a simple approach is to sequentially explore the temporal dimension and consider every possible activity period with $m$ time slots. This approach then chooses the corresponding optimal SGQ attendees for each activity period, and identifies the period with the smallest total social distance. However, this approach requires solving numerous SGQ problems and thereby is computationally intensive. In contrast, in this paper, we propose an algorithm, called *STGSelect*, to solve the problem. In addition to the strategies for *SGSelect*, we first identify pivot time slots, which are the time slots only required to be explored in the temporal dimension to reduce the running time. Moreover, we propose the availability pruning strategy for Algorithm STGSelect, and the strategy is able to leverage the correlation of the available time slots among candidate attendees, to avoid exploring an activity period eventually with fewer than $p$ available attendees.

The contributions of this paper are summarized as follows.

- We formulate two new and useful queries for an activity initiator, namely, SGQ and STGQ, to obtain the optimal set of attendees and a suitable activity time. SGQ can be used to plan for various types of activities by specifying the social radius constraint $s$ and the acquaintance constraint $k$, while STGQ also takes into account the temporal constraint. We prove that these two problems are NP-hard.

- We design Algorithm SGSelect and STGSelect to find the optimal solutions to SGQ and STGQ, respectively. We derive an Integer Programming optimization model for STGQ, which can also support SGQ. Moreover, we devise various strategies, including access ordering, distance pruning, acquaintance pruning, pivot time slots, and availability pruning, to prune unnecessary search space and obtain the optimal solution in smaller time. Our research result can be adopted in social networking websites and web collaboration tools as a value-added service.

The rest of this paper is summarized as follows. In Section 2, we introduce the related works. Section 3 formulates SGQ and explains the details of Algorithm SGSelect. Section 4 extends our study on SGQ to the more complex STGQ. The details of Algorithm STGSelect are also included. We present the experimental results in Section 5 and conclude this paper in Section 6.

## 2. RELATED WORKS

There is a tremendous need for activity planning. Even though some web applications, e.g., Meetup [6], have been developed to support activity coordination, these applications still require users to manually assign the activity time and participants. For example, using Meetup [6] or the Events function in Facebook [1], an activity initiator can specify an activity time and manually select some friends to send invitations to. In response, these friends inform the activity initiator whether they are able to attend or not. Obviously, the above-described manual activity coordination process is tedious and time-consuming. In contrast, the proposed STGQ, complementary to the above web applications, is able to automatically find a group of close friends to get together at a suitable activity time.

There are some related works on group formation (e.g., [10]), team formation (e.g., [14, 15]), and community search (e.g., [20]), but they focus on different scenarios and objectives. In addition, none of the above works consider the temporal dimension, i.e., schedules of attendees, to find a suitable activity time. Specifically, the work [10] uses real data to analyze the long-term evolution of the group formation in global social networks. On the contrary, we concern the social neighborhood of the activity initiator (i.e., her egocentric social network), since our goal is to find an optimal group of mutual acquaintances for the initiator. The studies on team formation [14, 15] find a group of experts covering all required skills, and minimize the communications cost between these experts. In contrast, we aim to find a group of friends satisfying social constraints and having common available time slots. The work [20] considers an unweighted social graph and studies the community search problem. The problem is to find a compact community that contains particular members, with the objective such as minimizing the total degree in the community. However, we consider a weighted social graph and minimize the total social distance within the group. Moreover, our problem includes an acquaintance constraint to ensure that each attendee can have at most $k$ other attendees unacquainted. This constraint is important because it allows us to adapt to the nature activity. In addition, the work [20] only provides an upper bound on the community size, instead of the exact number. Thus, users thereby may obtain a solution with an undesirably small number of attendees.

By minimizing the total social distance among the attendees, we are actually forming a cohesive subgroup in the social network. In the field of social network analysis, researches on finding various kinds of cohesive subgroups such as clique, n-club [17], and k-plex [19] have been conducted. However, most previous studies focus on finding all maximal cohesive subgroups or the maximum cohe-sive subgroups, but the total social distance is not considered. For example, finding the maximum k-plexes is discussed in [11, 16, 18], and enumeration of all maximal k-plexes is discussed in [21]. Instead of finding these extreme cases, we allow the activity initiator to specify an arbitrary group size. Note that the maximal subgroups or the maximum subgroups in the egocentric network of the initiator may not qualify the specified size. Hence, these existing works are not applicable to the SGQ problem. Besides, to the best of our knowledge, none of previous works on cohesive subgroups have considered the schedule of group members. Therefore, the STGQ problem is not addressed previously.

## 3. SOCIAL GROUP QUERY

In this section, we present the fundamental problem, namely Social Group Query (SGQ), which focuses on finding the optimal group of attendees satisfying the social radius constraint and the acquaintance constraint. In the following, we first present the problem formulation and research challenges in Section 3.1, and then prove that SGQ is NP-hard. Afterward, we propose an algorithm that can effectively prune unnecessary search space to obtain the optimal solution in Section 3.2.

### 3.1 Problem Definition

Given an activity initiator $q$ and her social graph $G = (V, E)$, where each vertex $v$ is a candidate attendee, and the distance on each edge $e_{u,v}$ connecting vertices $u$ and $v$ represents their social closeness. A social group query $SGQ(p, s, k)$, where $p$ is an activity size, $s$ is a social radius constraint, and $k$ is an acquaintance constraint, finds a set $F$ of $p$ vertices from $G$ to minimize the total social distance between $q$ and every vertex $v$ in $F$, i.e., $\sum_{v \in F} d_{v,q}$, where $d_{v,q}$ is the length of the minimum-distance path between $v$ and $q$ with at most $s$ edges, such that each vertex in $F$ is allowed to share no edge with at most $k$ other vertices in $F$.

The initiator can specify different $s$ for different kinds of activities. For example, the initiator can ensure that all invited attendees are directly acquainted with her by specifying $s = 1$. On the other hand, for a party, the initiator can specify a larger $s$ such that some friends of friends can also be invited to the party. Moreover, the initiator can vary $k$ for different kinds of activities, e.g., a small $k$ for a gathering where all the attendees know each other very well, while a larger $k$ for a more diverse event. Note that the size of solution space, i.e., the number of candidate groups, rapidly grows when $p$ and $s$ increase. On the other hand, $k$ serves as a filter to determine whether each candidate group satisfies the acquaintance constraint. Indeed, processing SGQ is an NP-hard problem (see Appendix B.1 for a formal proof). Fortunately, while the problem is very challenging, it is still tractable when the size of $s$ and $p$ are reasonable. In Section 3.2, we design an efficient query processing algorithm by pruning unqualified candidate groups.

THEOREM 1. *SGQ is NP-hard.*

PROOF. We prove the theorem in Appendix B.1. □

### 3.2 Algorithm Design

In this section, we propose a novel algorithm, namely SGSelect, to solve SGQ efficiently. Our idea is to first derive a *feasible graph* $G_F = (V_F, E_F)$ from $G$ based on our observation on the social radius constraint, such that there exists a path with at most $s$ edges from $q$ to each vertex in $G_F$. Starting from $q$, we iteratively explore $G_F$ to derive the optimal solution. At each iteration, we keep track of the set of vertices that satisfy the acquaintance constraint as the intermediate solution obtained so far (denoted as $V_S$). Initially, we set $V_S = \{q\}$, and let $V_A$ denote the set of remaining vertices in $V_F$, i.e., $V_A = V_F - V_S$. We select a vertex in $V_A$ and examine whether it is feasible (i.e., following the acquaintance constraint) to move this vertex to $V_S$ at each iteration, until $V_S$ has $p$ vertices and the process stops.

The selection of a vertex from $V_A$ at each iteration is critical to the performance of query processing. It is essential to avoid choosing a vertex $v$ that may significantly increase the total social distance or lead to violation of the acquaintance constraint. Based on our analysis on various properties and constraints in SGQ, we observe that the access order of nodes in constructing candidate groups is a key factor to the overall performance. It is important to take a priority to consider nodes that are very likely to be included in the final answer group, i.e., the optimal solution, which may facilitate effective early pruning of unqualified solutions. Additionally, social radius and acquaintance constraints can be exploited to facilitate efficient pruning of vertices which would not lead to the eventual answer. We summarize our ideas as follows.

**Access ordering.** To guide an efficient exploration of the solution space, we access vertices in $V_A$ following an order that incorporates (1) the increment of the total social distance and (2) the feasibility for the acquaintance constraint. Accordingly, we define the notion of *interior unfamiliarity* and *exterior expansibility* of $V_S$ to test the feasibility of examined vertices to the acquaintance constraint during the vertex selection.

**Distance pruning.** To avoid exploring vertices in $V_A$ that do not lead to a better solution in terms of total social distance, Algorithm SGSelect keeps track of the best feasible solution obtained so far and leverages its total social distance to prune unnecessary examinations of certain search space.

**Acquaintance pruning.** We explore the properties of the acquaintance constraint to facilitate search space pruning. Specifically, we define the notion of *inner degree* of the vertices in $V_A$ and derive its lower bound, such that a feasible solution can be derived from vertices in $V_S$ and $V_A$. The lower bound is designed to detect the stop condition when there exists no feasible solution after including any vertex in $V_A$.

To find the optimal solution, Algorithm SGSelect may incur an exponential time in query processing because SGQ is NP-hard. In the worst case, all candidate groups may need to be considered. However, by employing the above pruning strategies, the average running time of Algorithm SGSelect can be effectively reduced, as to be shown in Section 5. In the following, we present the details of the proposed algorithm.

### 3.2.1 Radius Graph Extraction

Obviously, the social radius constraint can effectively prune unnecessary candidates in the social network of the activity initiator. Thus, Algorithm SGSelect first extracts the vertices that satisfy the social radius constraint. A simple approach to ensure that the social radius constraint holds is to find the *minimum-edge path* (i.e., the shortest path with the minimum number of edges) between $q$ and every other vertex, and then remove those vertices that have their minimum-edge paths longer than $s$ edges. Nevertheless, the minimum-distance path with at most $s$ edges and the minimum-edge path can be different. As a result, the total distance of the minimum-edge path may not be the minimum distance. Moreover, the minimum-distance path may consist of more than $s$ edges which does not satisfy the social radius constraint. To address the above problem, we define the notion of *i-edge minimum distance*, which represents the total distance of the minimum-distance path with no more than $i$ edges as follows.

*Definition 1.* The $i$-edge minimum distance between the vertex $v$ and the vertex $q$ is $d_{v,q}^i = \min_{u \in N_v} \{d_{v,q}^{i-1}, d_{u,q}^{i-1} + c_{u,v}\}$, where $N_v$ is the set of neighboring vertices of $v$.

Based on dynamic programming, Algorithm SGSelect computes the $i$-edge minimum distance between the vertex $v$ and the vertex $q$ by iteratively deriving $d_{v,q}^i$ in terms of $d_{u,q}^{i-1}$ of each neighboring vertex $u$, for $1 \leq i \leq s$. Initially, we set $d_{v,q}^0$ as $\infty$ for every vertex $v$, $v \neq q$. We set $d_{q,q}^0$ as 0 and derive $d_{v,q}^1$ for every vertex $v$ in $N_q$. At the next iteration, we update $d_{v,q}^2$ for $v$ if there exists a neighbor $u$ of $v$ such that $d_{u,q}^1 + c_{u,v}$ is smaller than $d_{v,q}^1$. This case indicates that there is an alternate path from $v$ to $q$ via a neighbor $u$, and the path has a smaller total distance. Our algorithm repeats the above iterations with at most $s$ times for each vertex. Therefore, each vertex $v$ with $d_{v,q}^s < \infty$ is extracted in our algorithm to construct a feasible graph $G_F = (V_F, E_F)$. In the graph, the total distance of the minimum-distance path with at most $s$ edges (i.e., $d_{v,q}^s$) is adopted as the social distance between $v$ and $q$ (i.e., $d_{v,q}$). In other words, we can ensure that every vertex in $V_F$ satisfies the social radius constraint. Therefore, we consider $G_F$ in evaluating the SGQ for the rest of this paper.

### 3.2.2 Access Ordering

After constructing the feasible graph $G_F$, Algorithm SGSelect iteratively explores $G_F$ to find the optimal solution. Initially, the intermediate solution set $V_S$ includes only $q$, and the remaining vertex set $V_A$ is $V_F - \{q\}$. At each iteration afterward, we select and move a vertex from $V_A$ to $V_S$ in order to expand the intermediate solution in $V_S$. Therefore, $V_S$ represents a feasible solution when $|V_S| = p$, and the vertices in $V_S$ satisfy the acquaintance constraint. Next, our algorithm improves the feasible solution by backtracking the above exploration procedure to previous iterations and choosing an alternative vertex in $V_A$ to expand $V_S$. A branch-and-bound tree is maintained to record the exploration history for backtracking. This process continues until $V_S$ has $p$ vertices.

To reduce the running time and search space, the selection of a vertex at each iteration is critical. Naturally, we would like to include a vertex that minimizes the increment of the total social distance. Nevertheless, the connectivity of the selected vertex imposes additional requirements for satisfying the acquaintance constraint. Thus, we introduce the notion of *interior unfamiliarity* and *exterior expansibility* with respect to the intermediate solution set $V_S$ to exploit the acquaintance constraint in query processing.

*Definition 2.* The interior unfamiliarity of $V_S$ is $U(V_S) = \max_{v \in V_S} |V_S - \{v\} - N_v|$, where $N_v$ is the set of neighboring vertices of $v$ in $G_F$. The set $V_S - \{v\} - N_v$ refers to the set of non-neighboring vertices of $v$ in $V_S$.

As shown later, the interior unfamiliarity of possible intermediate solution sets are taken into account in deciding which vertex is to be included in the process of generating the candidate groups. It is preferable to first include a well-connected vertex that results in the intermediate solution set with low interior unfamiliarity since it may make selections of other vertices in the later iterations easier. Next, we define the exterior expansibility of an intermediate solution set $V_S$, denoted by $A(V_S)$, as the maximum number of vertices that $V_S$ can be expanded from.

*Definition 3.* The exterior expansibility of $V_S$ is $A(V_S) = \min_{v \in V_S} \{|V_A \cap N_v| + (k - |V_S - \{v\} - N_v|)\}$, where the first set (i.e., $V_A \cap N_v$) contains the neighboring vertices of $v$ in $V_A$ and the second set (i.e., $V_S - \{v\} - N_v$) contains the non-neighboring vertices of $v$ in $V_S$.

Since the number of existing non-neighboring vertices of $v$ in $V_S$ is $|V_S - \{v\} - N_v|$, we can select at most $k - |V_S - \{v\} - N_v|$ extra non-neighboring vertices of $v$ from $V_A$ to expand $V_S$; otherwise, vertex $v$ would have more than $k$ non-neighboring vertices in $V_S$ and violate the acquaintance constraint. Therefore, for a vertex $v$ in $V_S$, there are at most $|V_A \cap N_v|$ neighboring vertices and $k - |V_S - \{v\} - N_v|$ non-neighboring vertices to be selected from $V_A$ in order to expand $V_S$.

When selecting a vertex $v$ to expand $V_S$, we consider both the increment of the total social distance caused by $v$ and the connectivity of vertices in the new intermediate solution set containing $v$, which is captured by $U(V_S \cup \{v\})$ and $A(V_S \cup \{v\})$. Specifically, Algorithm SGSelect chooses the vertex $v$ with the minimum social distance to $q$ that satisfies the following two conditions for interior unfamiliarity and exterior expansibility, respectively.

**Interior Unfamiliarity Condition.** The first condition considers the interior unfamiliarity. Note that a small value of interior unfamiliarity indicates that every vertex $v \in V_S$ has plenty of neighboring vertices in $V_S$, i.e., the current intermediate solution set $V_S$ is likely to be expanded into feasible solutions satisfying the acquaintance constraint. Based on this observation, we employ the interior unfamiliarity condition, i.e., $U(V_S \cup \{v\}) \leq k \left[\frac{|V_S \cup \{v\}|}{p}\right]^\theta$, where $\theta \geq 0$ and $\frac{|V_S \cup \{v\}|}{p}$ is the proportion of attendees that have been considered, to ensure that the value of interior unfamiliarity remains small when a vertex $v$ is selected. Note that the right-hand-side (RHS) of the inequality reaches its maximum, i.e., $k$, when $\theta$ is fixed as 0. With $\theta = 0$, it is flexible to find a vertex $v$ with a small social distance. However, if a vertex $v$ resulting in $U(V_S \cup \{v\}) = k$ is selected, the vertex with $k$ non-neighboring vertices in the set $V_S \cup \{v\}$ is required to connect to all the vertices chosen from $V_A$ at later iterations. Thus, the feasibility of selecting other qualified vertices in later iterations is thereby decreased. In contrast, a larger $\theta$ allows SGSelect to choose a vertex from $V_A$ that connects to more vertices in $V_S$ to ensure the feasibility at later iterations. Note that the RHS of the condition increases when $V_S$ includes more vertices. On the other hand, the algorithm reduces $\theta$ if there exists no vertex in $V_A$ that can satisfy the above condition. When $\theta$ decreases to 0 and the above condition still does not hold for all vertices, i.e., $U(V_S \cup \{v\}) > k$, Algorithm SGSelect removes every $v$ from $V_A$, because adding $v$ to $V_S$ does not generate a feasible solution.

**Exterior Expansibility Condition.** Now we discuss the second condition based on the exterior expansibility, which represents the maximum number of vertices in $V_A$ that can be considered for expanding the intermediate solution set $V_S$, and this value must be no smaller than the number of attendees required to be added later, i.e., $p - |V_S|$. Therefore, SGSelect chooses the vertex $v$ from $V_A$ that can satisfy the exterior expansibility condition, i.e., $A(V_S \cup \{v\}) \geq (p - |V_S \cup \{v\}|)$. If the inequality does not hold, the new intermediate solution set obtained by adding $v$ is not expansible, as shown by the following lemma.

LEMMA 1. *Given that $A(V_S) < (p - |V_S|)$, there must exist at least one vertex $v$ in $V_S$ such that $v$ cannot follow the acquaintance constraint for every possible selection of vertices from $V_A$.*

PROOF. If $A(V_S) < (p - |V_S|)$, then we can find at least one vertex $v$ in $V_S$ such that $|V_A \cap N_v| + (k - |V_S - \{v\} - N_v|) < (p - |V_S|)$. In other words, $(k - |V_S - \{v\} - N_v|) < (p - |V_S|) - |V_A \cap N_v|$. As mentioned above, $|V_S - \{v\} - N_v|$ is the number of non-neighboring vertices for $v$, and $k - |V_S - \{v\} - N_v|$ thereby represents the "quota" for $v$ to choose non-neighbor ver-

tices from $V_A$. For any possible selection $\widehat{V}_A \subseteq V_A$, let $\widehat{\lambda}_A$ denote the number of neighbor vertex of $v$ in $\widehat{V}_A$. Since $\widehat{\lambda}_A \leq |V_A \cap N_v|$, $(p - |V_S|) - |V_A \cap N_v| \leq (p - |V_S|) - \widehat{\lambda}_A$. Therefore, if $A(V_S) < (p - |V_S|)$, then $(k - |V_S - \{v\} - N_v|) < (p - |V_S|) - \widehat{\lambda}_A$, and $v$ does not have enough quota to support $\widehat{V}_A$ for satisfying the acquaintance constraint. The lemma follows. □

### 3.2.3 Distance and Acquaintance Pruning

In the following, we further exploit two pruning strategies to reduce the search space. Our algorithm aims to obtain a feasible solution early since the total social distance of this solution can be used for pruning unnecessary candidates. At each iteration, the following distance pruning strategy avoids exploring the vertices in the remaining vertex set $V_A$ if they do not lead to a solution with a smaller total social distance.

LEMMA 2. *The distance pruning strategy stops selecting a vertex from $V_A$ to $V_S$ if $D - \sum_{v \in V_S} d_{v,q} < (p - |V_S|) \min_{v \in V_A} d_{v,q}$, where $D$ is the total social distance of the best feasible solution obtained so far. The distance pruning strategy can prune the search space with no better solution.*

PROOF. If the above condition holds, it is impossible to find an improved solution by exploring $V_A$, since the total social distance of any new solution must exceed $D$ when we select $p - |V_S|$ vertices from $V_A$. Algorithm SGSelect considers $\min_{v \in V_A} d_{v,q}$ in distance pruning to avoid sorting the distances of all vertices in $V_A$, which requires additional computation and may not be scalable for a large social network. Please note that as the best obtained solution improves at later iterations, we are able to derive a smaller upper bound in the LHS, and thus prune a larger search space with distance pruning. The lemma follows. □

In addition to pruning the search space that does not lead to a smaller total social distance, we also propose an acquaintance pruning strategy that considers the feasibility of selecting vertices from $V_A$, and stops exploring $V_A$ if there exists no solution that can satisfy the acquaintance constraint. Earlier in this section, the interior unfamiliarity and the exterior expansibility consider the connectivity between the vertices in only $V_S$, and the connectivity between the vertices in $V_S$ and $V_A$, respectively. Here the acquaintance pruning strategy focuses on the edges between the vertices in $V_A$. Note that all vertices in $V_A$ are excluded from expansion (and thus the corresponding $V_S$ is pruned) if $\sum_{v \in V_A} |V_A \cap N_v| < (p - |V_S|)(p - |V_S| - k)$ holds. The LHS of the above inequality is the total inner degree of all vertices in $V_A$, where the *inner degree* of a vertex in $V_A$ considers only the edges connecting to other vertices in $V_A$. The RHS is the lower bound on the total inner degree on any set of vertices extracted from $V_A$ to expand $V_S$ into a solution satisfying the acquaintance constraint. Specifically, our algorithm needs to select $p - |V_S|$ vertices from $V_A$, and hence the inner degree of any selected vertex cannot be smaller than $p - |V_S| - k$; otherwise, the vertex must be unacquainted with more than $k$ vertices and violate the acquaintance constraint.

The above strategy can be improved by replacing the LHS of the inequality with $\sum_{v \in M_A} |V_A \cap N_v|$, where $M_A$ denotes the set of $p - |V_S|$ vertices in $V_A$ with the largest inner degrees. Therefore, with $M_A$, our algorithm is able to stop the search earlier, and prune off more infeasible solutions because $M_A \subseteq V_A$, and $\sum_{v \in M_A} |V_A \cap N_v| \leq \sum_{v \in V_A} |V_A \cap N_v|$. However, to obtain $M_A$, sorting the vertices in $V_A$ according to their inner degrees requires additional computation. Therefore, our algorithm finds another value between $\sum_{v \in M_A} |V_A \cap N_v|$ and $\sum_{v \in V_A} |V_A \cap N_v|$, and the value can be obtained easily. Specifically, the acquaintance pruning is specified as follows.

LEMMA 3. *The acquaintance pruning strategy stops selecting a vertex from $V_A$ to $V_S$ if $\sum_{v \in V_A} |V_A \cap N_v| - (|V_A| - p + |V_S|) \min_{v \in V_A} |V_A \cap N_v| < (p - |V_S|)(p - |V_S| - k)$, and the acquaintance pruning strategy can prune the search space with no feasible solution.*

PROOF. To avoid the sorting in $\sum_{v \in M_A} |V_A \cap N_v|$, we consider only the vertex with the minimum inner degree $\min_{v \in V_A} |V_A \cap N_v|$, and the LHS is an upper bound on $\sum_{v \in M_A} |V_A \cap N_v|$. The reason is that $|V_A| - p + |V_S|$ is the number of vertices not extracted from $V_A$, and the second term of the LHS thereby represents a lower bound on the total inner degree of the vertices not extracted from $V_A$. Therefore, the LHS is an upper bound on the total inner degree of the vertices extracted from $V_A$. This upper bound can be employed to improve the acquaintance pruning strategy, since $\sum_{v \in M_A} |V_A \cap N_v| \leq \sum_{v \in V_A} |V_A \cap N_v| - (|V_A| - p + |V_S|) \min_{v \in V_A} |V_A \cap N_v| \leq \sum_{v \in V_A} |V_A \cap N_v|$. The lemma follows. □

In the following, we prove that our algorithm with the above strategies finds the optimal solution.

THEOREM 2. *SGSelect obtains the optimal solution to SGQ.*

PROOF. We prove the theorem in Appendix B.2. □

Please refer to Example 2 in Appendix A for illustration of Algorithm SGSelect, and the pseudo code of SGSelect is provided in Appendix C.

## 4. SOCIAL-TEMPORAL GROUP QUERY

In the following, we extend SGQ to STGQ by exploring the temporal dimension and formulate the problem in Section 4.1. STGQ is more complex than SGQ because there may exist numerous activity periods with different candidate groups. An intuitive approach is to first find the SGQ solution for each individual activity period and then select the one with the minimum total social distance. However, this approach is computationally expensive. To address this issue, in Section 4.2, we identify pivot time slots, the only time slots required to be explored in the temporal dimension, to facilitate efficient STGQ processing. Moreover, we propose the availability pruning strategy to leverage the correlation in the available time slots among candidate attendees to avoid exploring an unsuitable activity period.

### 4.1 Problem Definition

STGQ generalizes SGQ by considering the available time of each candidate attendee via the *availability constraint*, which ensures that all selected attendees are available in a period of $m$ time slots. Given an activity initiator $q$ and her social graph $G = (V, E)$, where each vertex $v$ is a candidate attendee, and the distance on each edge $e_{u,v}$ connecting vertices $u$ and $v$ represents their social closeness. A social-temporal group query $STGQ(p, s, k, m)$, where $p$ is an activity size, $s$ is a social radius constraint, $k$ is an acquaintance constraint, and $m$ is an activity length, finds a time slot $t$ and a set $F$ of $p$ vertices from $G$ to minimize the total social distance between $q$ and every vertex in $F$, i.e., $\sum_{u \in F} d_{u,q}$, where $d_{u,q}$ is the length of the minimum-distance path between $u$ and $q$ with at most $s$ edges, such that each vertex $u$ in $F$ is allowed to share no edge with at most $k$ other vertices in $F$, and $u$ is available from time slot $t$ to $t + m - 1$.

STGQ is also an NP-hard problem because STGQ can be reduced to SGQ if every candidate attendee is available in all time slots. In Section 4.2, we design an efficient query processing algorithm STGSelect by leveraging the idea of pivot time slots to effectively reduce the number of time slots to be processed. Additionally, we derive the Integer Programming formulation of STGQ (see Appendix D), which is included in Section 5 for comparison.

## 4.2 Algorithm Design

An intuitive approach to evaluate STGQ is to consider the social dimension and the temporal dimension separately, by sequentially exploring each time slot $t$ and the candidate attendees who are available from $t$ to $t + m - 1$ (i.e., $m$ consecutive time slots). However, the running time significantly grows when the number of time slots increases. Therefore, we devise Algorithm STGSelect, which explores the following features in the temporal dimension to reduce search space and running time.

**Pivot time slot.** We consider only a limited number of slots, namely, the *pivot time slots*, to find the solution. STGSelect returns optimal solutions even though only parts of the slots are considered.

**Access ordering.** In addition to interior unfamiliarity and exterior expansibility discussed earlier, we further consider the solution quality and the feasibility based on the availability constraint. Algorithm STGSelect constructs the $V_S$ with vertices which have more available time slots in common to find an initial feasible solution and then chooses the vertices in $V_A$ with smaller social distances to improve the solution.

**Availability pruning.** In addition to the distance and acquaintance pruning discussed in Section 3.2, we propose the availability pruning strategy to stop the algorithm when selecting any vertex from $V_A$ never leads to a solution with $m$ available time slots.

To find the optimal solution, STGSelect is expected to have an exponential-time complexity because STGQ is NP-hard. In the worst case, all candidate groups in all time slots may need to be considered. However, as shown in Section 5, the average running time of the proposed algorithm with the above strategies can be effectively reduced, especially for a large $m$. In the following, we describe the details of Algorithm STGSelect, paying special attention on the temporal dimension. Instead of considering the interval from $t$ to $t + m - 1$ for each time slot $t$, our algorithm leverages the pivot time slots defined as follows to reduce running time.

LEMMA 4. *A time slot is a pivot time slot if the ID of the slot is $im$, where $i$ is a positive integer. Any feasible solution to STGQ must include exactly one pivot time slot.*

PROOF. If a solution does not span over a pivot time slot, the solution must have fewer than $m$ slots because there are $m-1$ time slots between any two consecutive pivot time slots. If a solution contains more than one pivot time slot, the solution includes more than $m$ slots, and the above two cases are not feasible. Moreover, there must exist an integer $i^*$ such that the optimal solution resides in an interval starting from slot $(i^* - 1)m + 1$ to $(i^* + 1)m - 1$, corresponding to pivot time slot $i^*m$. If the optimal solution is not located in the above interval, the optimal solution must include at least two pivot time slots and thereby is infeasible, or the optimal solution must reside in the corresponding interval for pivot time slot $(i^* - 1)m$ or $(i^* + 1)m$. The lemma follows. □

*Definition 4.* Every vertex $v$ in the feasible graph $G_F^{im} = (V_F^{im}, E_F^{im})$ for pivot time slot $im$ has at least $m$ consecutive available time slots in the interval from slot $(i-1)m+1$ to $(i+1)m-1$. Moreover, there exists a path from $q$ to $v$ with at most $s$ edges.

For each pivot time slot $im$, Algorithm STGSelect extends SGSelect by considering the temporal information when selecting a vertex from $V_A$ to $V_S$. Specifically, let $T_S$ denote the set of consecutive time slots available for all vertices in $V_S$, and $T_S$ must contain slot $im$. In other words, $T_S$ will be a feasible solution to the STGQ when $V_S$ includes $p$ vertices satisfying the acquaintance constraint, and $|T_S| \geq m$. At each iteration, for each vertex in $V_A$, Algorithm STGSelect considers the social distance to $q$ during the selection to reduce the objective value. However, we also consider the temporal availability of the vertex to avoid choosing a vertex that leads to a small increment of the total social distance but ends up with redundant examination of solutions eventually disqualified by the availability constraint. In other words, in addition to interior unfamiliarity and exterior expansibility as described in Section 3.2, we define the notion of *temporal extensibility* as follows.

*Definition 5.* The temporal extensibility of $V_S$ is $X(V_S) = |T_S| - m$. A larger temporal extensibility ensures that many vertices in $V_A$ with good quality in the temporal dimension can be selected by our algorithm afterward.

**Temporal Extensibility Condition.** To consider both the solution quality and feasibility in the temporal dimension, Algorithm STGSelect chooses the vertex $u$ with the minimum social distance to $q$, and $u$ must satisfy $X(V_S \cup \{u\}) \geq (m-1) \left[ \frac{p - |V_S \cup \{u\}|}{p} \right]^\phi$, where $\phi \geq 1$ and $\frac{p - |V_S \cup \{u\}|}{p}$ is the proportion of attendees that have not been considered. The RHS grows when $\phi$ decreases, and the above condition enforces that the result $V_S \cup \{u\}$ must be more temporal extensible, i.e., more available time slots are shared by all vertices in the result, and hence more vertices in $V_A$ are eligible to be selected at later iterations. In the extreme case, if $\phi = 1$, the above condition requires that the result contains almost $2m - 1$ available time slots when $V_S = \{q\}$, because the RHS is close to $m - 1$. In contrast, as $\phi$ grows, our algorithm is able to choose a vertex with a smaller social distance because more vertices can satisfy the above condition. Please note that $\phi$ is increased by the algorithm if there exists no vertex in $V_A$ that can satisfy the above condition, and the RHS approaches 0 in this case. For the case that leads to $X(V_S \cup \{u\}) < 0$, we remove $u$ from $V_A$ because adding $u$ to $V_S$ results in unqualified solutions that are infeasible in the temporal dimension.

In addition to distance pruning and acquaintance pruning that consider the social dimension, we propose availability pruning in the temporal dimension. The strategy enables our algorithm to stop exploring $V_A$ if there exists no solution that can satisfy the availability constraint. The above temporal extensibility considers the available time slots for vertices in $V_S$. In contrast, availability pruning reduces the search space according to the available time slots of vertices in $V_A$. Specifically, for each pivot time slot $im$, let $\bar{t}_A^+$ and $\bar{t}_A^-$ denote the time slots closest to $im$, such that all vertices in $V_A$ are not available in the two time slots, where $\bar{t}_A^+ > im$ and $\bar{t}_A^- < im$, respectively. Therefore, we are able to stop considering $V_A$ when $\bar{t}_A^+ - \bar{t}_A^- \leq m$. In this case, the solution is infeasible since the interval starting from $\bar{t}_A^- + 1$ to $\bar{t}_A^+ - 1$ contains fewer than $m$ time slots. This strategy can be further improved by considering the number of vertices that are not available for each time slot, and the availability pruning strategy is formally specified as follows.

LEMMA 5. *The availability pruning strategy stops selecting a vertex from $V_A$ to $V_S$ if $\bar{t}_A^+(|V_A| - p + |V_S| + 1) - \bar{t}_A^-(|V_A| - p + |V_S| + 1) \leq m$, where $\bar{t}_A^+(n)$ and $\bar{t}_A^-(n)$ denote the time slots closest to $im$, such that at least $n$ vertices in $V_A$ are not available,*

and $\bar{t}_A^+(n) > im$ and $\bar{t}_A^-(n) < im$, respectively. Moreover, the availability pruning strategy can prune the search space with no feasible solution.

PROOF. If the above condition holds, there are at most $p - |V_S| - 1$ vertices of $V_A$ available in each of the above two slots, and we can never find a feasible solution because Algorithm STGSelect is required to choose $p - |V_S|$ vertices from $V_A$ for a common available interval with at least $m$ time slots. The lemma follows. □

THEOREM 3. *STGSelect obtains the optimal solution to STGQ.*

PROOF. We prove the theorem in Appendix B.3. □

Please refer to Example 3 in Appendix A for illustration of Algorithm STGSelect, and the pseudo code of STGSelect is provided in Appendix C.

## 5. EXPERIMENTAL RESULTS

In this section, we evaluate the performance and analyze the solution quality of the proposed algorithms. We perform a series of sensitivity tests to study the impact of query parameters with both real and synthetic datasets. In the following, Section 5.1 describes the experiment setup. Section 5.2 then presents the experimental results on performance and solution quality.

### 5.1 Experiment Setup

We invite 194 people from various communities, e.g., schools, government, business, and industry to join our experiment, and Google Calendar [3] is utilized to collect the schedules from them. The social distance of each edge is derived according to the interaction between the two corresponding people [10, 12, 13], such as the frequency of meeting, phone calls, and mails. Moreover, a synthetic dataset with 12800 people is generated from a coauthorship network [7], while the schedule of each person in each day is randomly assigned from the above 194-people real dataset. The proposed algorithms are implemented in an IBM 3650 server with two Quadcore Intel X5450 3.00 GHz CPUs and 8 GB RAM.

To evaluate the solution quality, we compare STGSelect with PCArrange, which is an algorithm imitating the behavior of manual coordination via phone calls, where the initiator $q$ sequentially invites close friends first and then finds out the common available time slots. Note that PCArrange does not include the acquaintance constraint, but we can obtain a parameter $k_h$ from each result, such that every attendee has at most $k_h$ other attendees unacquainted. In other words, $k_h$ is the observed $k$ for each activity in PCArrange. In addition to STGSelect, we compare PCArrange with an algorithm STGArrange, which is designed to utilize STGSelect for finding the smallest $k$ that can achieve the total social distance no more than the one in PCArrange. In other words, starting from $k = 0$, STGArrange incrementally increases $k$ in STGSelect to find the first solution no worse than the one in PCArrange, in order to evaluate $k$.

### 5.2 Analysis of Experimental Results

For SGQ, we compare Algorithm SGSelect with two other approaches: the baseline algorithm for SGQ, i.e., considering all possible candidate groups, and the Integer Programming model for SGQ in Appendix D with CPLEX [4], which is an integer programming optimizer that can support parallel computation. We first compare the running time of Algorithm SGSelect and the baseline algorithm for SGQ with different numbers of attendees, i.e., $p$, in Figure 1(a). Due to the space constraint, we only include the experimental results with $k = 2$ and $s = 1$, and the trends for other parameter settings, such as $s = 2$, are identical. The results indicate that Algorithm SGSelect outperforms the baseline algorithm for SGQ, and the difference between them becomes more significant as $p$ grows. This is because the baseline algorithm for SGQ considers numerous candidate groups, and the processing effort of each candidate group increases with $p$, while Algorithm SGSelect effectively prunes the solution space with the proposed access ordering, distance pruning, and acquaintance pruning strategies. Integer programming requires more time than the baseline algorithm for SGQ because it is a general-purpose optimizer. However, the running time grows slower because it can support parallel computation to fully utilize the 8-core processing power in IBM 3650. In contrast, both SGSelect and the baseline algorithm are single-thread algorithms, and the results show that SGSelect is much faster than integer programming for finding the optimal solutions.

Figure 1(b) shows the results with different social radius constraints, i.e., $s$. The reason that Algorithm SGSelect is able to effectively reduce the running time is that it first derives a much smaller feasible graph $G_F$ by exploiting the social radius constraint and then prunes unnecessary branches. Besides the social radius constraint, we also compare the running time of these two approaches under different acquaintance constraints, i.e., $k$. As shown in Figure 1(c), different $k$ only slightly change the running time. This is because the value of $k$ does not directly affect the number of candidate groups required to be considered. In addition, the result shows that Algorithm SGSelect outperforms the baseline algorithm by near two orders under every value of $k$. We also compare the algorithms with different network sizes as shown in Figure 1(d), and we can see that the running time of Algorithm SGSelect is still much smaller than the time of the baseline algorithm.

To evaluate the performance on STGQ, we compare our algorithm STGSelect with the baseline algorithm for STGQ, i.e., sequentially considering each time slot and solving the corresponding SGQ problem. In Figure 1(e), we report the running time of these two algorithms under different activity lengths, i.e., $m$. The results show that Algorithm STGSelect significantly outperforms the baseline algorithm, especially when $m$ increases. In contrast, Figure 1(f) presents the running time of Algorithm STGSelect and the baseline algorithm with different lengths of schedules provided by users. More time slots need to be considered in a longer schedule. As shown in these two figures, Algorithm STGSelect yields much less running time, and the difference increases as $m$ or the schedule length increases. The reason is that with the idea of pivot time slots, Algorithm STGSelect can concurrently consider $m$ candidate activity periods spanning the same pivot time slot, while the baseline algorithm considers these candidate periods separately.

Figure 1(g) and Figure 1(h) compare $k$ and the total social distance obtained by STGArrange and PCArrange with different $p$. For each $p$, a smaller $k$ corresponds to a better solution since each attendee is acquainted with more people in the group. Similarly, a smaller total social distance represents better solution quality since the attendees are more familiar with the activity initiator. These two figures indicate that the quality of the solutions obtained by STGArrange outperforms the ones obtained by PCArrange. In other words, STGArrange is able to find a solution with a much smaller $k$, while the total social distance is also smaller than the one in PCArrange.

## 6. CONCLUSION

To the best of our knowledge, there is no real system and existing work in the literature that addresses the issues of automatic activity planning based on social and temporal relationship of an initiator and activity attendees. In this paper, we define two new

403

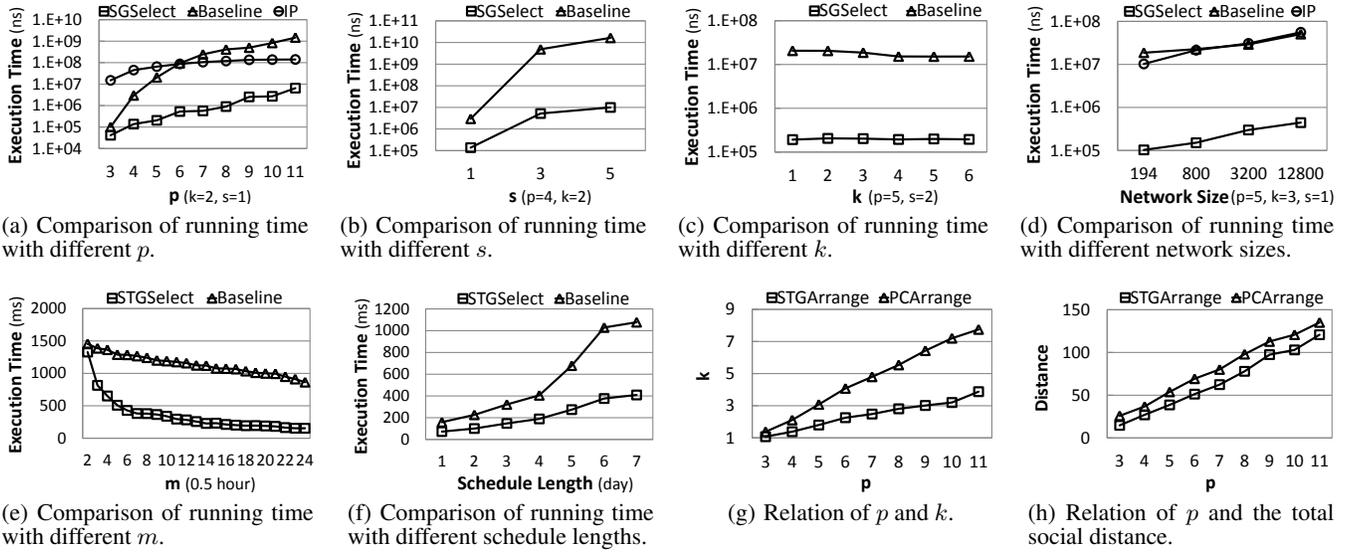

Figure 1: Experimental results of SGQ and STGQ.

and useful queries, namely, SGQ and STGQ, to obtain the optimal set of attendees and a suitable activity time. We devise two algorithms, namely, SGSelect and STGSelect, to find the optimal solutions to SGQ and STGQ, respectively. We show that the problem is NP-hard and propose an Integer Programming model for STGQ, which can also support SGQ. Various strategies, including access ordering, distance pruning, acquaintance pruning, pivot time slots, and availability pruning are explored to prune unnecessary search space and obtain the optimal solution efficiently. Experimental results indicate that the proposed algorithm significantly outperforms the algorithm that represents manual coordination by the initiator. Our research result can be adopted in social networking websites and web collaboration tools as a value-added service, and we are now implementing the proposed algorithms in Facebook.

## 7. ACKNOWLEDGEMENTS

This paper was supported in part by National Science Council of Taiwan under Contract NSC99-2219-E-001-001.

# APPENDIX
## A. ILLUSTRATING EXAMPLES

EXAMPLE 1. Consider a toy example, as illustrated in Figure 2, where Casey Affleck would like to invite some friends to discuss a new movie in which he is going to play. Figure 2(a) shows a possible social network of Casey Affleck, constructed based on the cooperation relationship extracted from Yahoo! Movies [9]. Assume that Casey Affleck is trying to find three mutually acquainted friends. Issuing an SGQ with $p = 4$ and $s = 1$, which returns {George Clooney, Robert De Niro, Casey Affleck, Michelle Monaghan}, does not work since the three close friends of Casey Affleck are not acquainted to each other (as shown in Figure 2(a)). Instead, by issuing an SGQ with $p = 4$, $s = 1$, and $k = 0$, a better list of invitees {George Clooney, Brad Pitt, Julia Roberts, Casey Affleck}, where everyone knows each other very well, is obtained.

To answer this query, all directly connected friends of Casey Affleck, together with Casey Affleck himself, i.e., {George Clooney, Robert De Niro, Brad Pitt, Julia Roberts, Casey Affleck, Michelle Monaghan}, are candidates. Thus, an intuitive approach to answer this SGQ is to first enumerate ten (i.e., $C_{4-1}^{6-1} = 10$) possible four-person candidate groups including Casey Affleck himself. Figure 2(b) illustrates the enumeration process of candidate groups, in accordance with the increasing order of user IDs. Note that the numbers 64 and 65 indicate the total social distances of qualified candidate groups. These ten non-duplicate candidate groups constitute the whole solution space, from which we eliminate the ones disqualified by the acquaintance constraint. Finally, we obtain the final answer {George Clooney, Brad Pitt, Julia Roberts, Casey Affleck}, i.e., the group with the smallest total social distance among qualified candidate groups.

Note that the initiator may also loosen the social constraints (i.e., $s > 1$ and $k > 0$) to consider more candidates. Assume that Casey Affleck issues another SGQ, inviting five friends taking the six-seat chartered plane to visit the refugee children in Haiti with him. Now Casey Affleck issues an SGQ with $p = 6$, $s = 2$ and $k = 2$, and the number of candidates is enlarged to eight (and the number of candidate groups is enlarged to $C_{6-1}^{8-1} = 21$) since the friends of his friends can also be considered. The optimal result for this SGQ is {Angelina Jolie, George Clooney, Robert De Niro, Brad Pitt, Julia Roberts, Casey Affleck}, and we can see that this is a tight group satisfying the social constraints in the sample social network.

However, Casey Affleck finds out that these six attendees have no available time in common when he expects the length of activity time as 3, i.e., three consecutive time slots. Figure 2(c) shows the schedule of candidates, with their available time slots marked by circles. Therefore, he turns to issue an STGQ with $p = 6$, $s = 2$, $k = 2$ and $m = 3$. An intuitive approach to find the answer is to issue an SGQ independently for each activity period of length 3. In this example, the time interval $[ts_1, ts_6]$ can be divided into four candidate activity periods, $[ts_1, ts_3]$, $[ts_2, ts_4]$, $[ts_3, ts_5]$ and $[ts_4, ts_6]$. For each activity period, the candidate attendees need to be available for all the time slots in this activity period in order to be considered.

After issuing four SGQs, the obtained result is {Angelina Jolie, George Clooney, Robert De Niro, Brad Pitt, Matt Damon, Casey Affleck} for both of $[ts_2, ts_4]$ and $[ts_3, ts_5]$. Therefore, the intuitive approach returns the group {Angelina Jolie, George Clooney, Robert De Niro, Brad Pitt, Matt Damon, Casey Affleck} and the activity period $[ts_2, ts_4]$ (or $[ts_3, ts_5]$) as the optimal solution. □

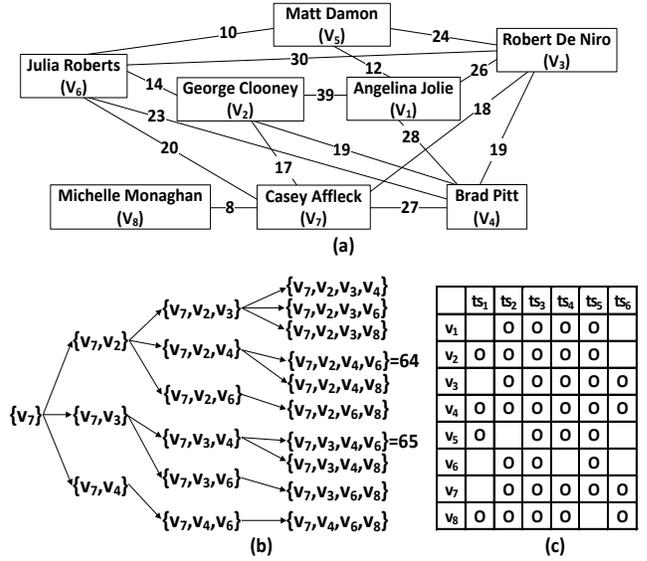

Figure 2: An illustrative example for SGQ and STGQ. (a) The sample social network, (b) the dendrogram of candidate group enumeration and (c) schedules of candidate attendees.

EXAMPLE 2. In this illustrating example for SGSelect, we revisit the social network in Figure 2(a) and assume that $v_7$ is the initiator issuing an SGQ with $p = 4$, $s = 1$, and $k = 1$. All candidate attendees $v_i$ with $d_{v_i,v_7}^1 < \infty$ are extracted and placed in Figure 3(a), with their social distances to $v_7$ listed in Figure 3(b).[3] In the beginning, $V_S = \{v_7\}$ and $V_A = \{v_2, v_3, v_4, v_6, v_8\}$. We first consider selecting $v_2$ (i.e., the vertex with the smallest social distance) from $V_A$ to expand $V_S$. Afterward, we have $A(V_S \cup \{v_2\}) = 3^4$ and $(p - |V_S \cup \{v_2\}|) = 4 - 2 = 2$, which means that the exterior expansibility condition holds if we select $v_2$. In addition, we find $U(V_S \cup \{v_2\}) = 0^5$ and $k\left[\frac{|V_S \cup \{v_2\}|}{p}\right]^\theta = 1 \times (\frac{2}{4})^2 = \frac{1}{4}$ (assume $\theta = 2$), which means the interior unfamiliarity condition also holds, and hence $v_2$ is selected. Now we have $V_S = \{v_2, v_7\}$ and $V_A = \{v_3, v_4, v_6, v_8\}$, and the next vertex to be considered is $v_3$ according to the social distance. The exterior expansibility condition holds when $v_3$ is selected, since $A(V_S \cup \{v_3\}) = 1 \geq (p - |V_S \cup \{v_3\}|) = 1$; however, it violates the interior unfamiliarity condition, since $U(V_S \cup \{v_3\}) = 1 > k\left[\frac{|V_S \cup \{v_3\}|}{p}\right]^\theta = 1 \times (\frac{3}{4})^2 = \frac{9}{16}$. We do not reduce the $\theta$ here because there are still more vertices in $V_A$, and we put $v_3$ in parenthesis and temporarily skip it. The current $V_A$ is $\{(v_3), v_4, v_6, v_8\}$, and the next vertex to be considered is $v_6$. We select $v_6$ since both of the exterior expansibility condition and the interior unfamiliarity condition hold, and then we have $V_S = \{v_2, v_6, v_7\}$ and $V_A = \{v_3, v_4, v_8\}$. Again, selecting $v_3$ violates the interior unfamiliarity condition, and we temporarily skip $v_3$. When selecting $v_8$, we find out that it violates the exterior expansibility condition and then remove $v_8$ from $V_A$. Therefore, we choose $v_4$ instead and obtain the first feasible solution $\{v_2, v_4, v_6, v_7\}$ (total social distance = 64). Note that if we set a small $\theta$ and allow $v_3$ to be selected earlier, it leads to the generation of an infeasible candidate group $\{v_2, v_3, v_6, v_7\}$, instead of the first feasible solution. If we can obtain the first feasible solution early, we are able to leverage the distance pruning strategy. Now we backtrack one step to the state $V_S = \{v_2, v_6, v_7\}$ and $V_A = \{(v_3)\}$, and we reduce $\theta$ since there is no other vertex in $V_A$. However, selecting $v_3$ still violates the interior unfamiliarity condition when we reduce $\theta$ to 0, since $U(V_S \cup \{v_3\}) = 2$ and $k \left[ \frac{|V_S \cup \{v_3\}|}{p} \right]^\theta = 1 \times (\frac{4}{4})^0 = 1$. Therefore, we can remove $v_3$ and backtrack to the state $V_S = \{v_2, v_7\}$ and $V_A = \{(v_3), v_4, v_8\}$. In this branch, we first remove $v_8$ due to the violation of the exterior expansibility condition, and later we select $v_4$ and $v_3$ to obtain the second feasible solution $\{v_2, v_3, v_4, v_7\}$ (total social distance = 62). Note that when reducing $\theta$ to 0 here, the interior unfamiliarity condition holds and we can select $v_3$, since $U(V_S \cup \{v_3\}) = 1$ and $k \left[ \frac{|V_S \cup \{v_3\}|}{p} \right]^\theta = 1 \times (\frac{4}{4})^0 = 1$. Now we further backtrack to the state $V_S = \{v_7\}$ and $V_A = \{v_3, v_4, v_6, v_8\}$[6] and select $v_3$. However, since $62 - 18 < (4 - 2) \times 23$, the distance pruning strategy then stops selecting a vertex from $V_A$. Later, we find out that there is no need to explore the last state $V_S = \{v_7\}$ and $V_A = \{v_4, v_6, v_8\}$, since $(1 + 1 + 0) - (3 - 4 + 1) \times 0 < (4 - 1) \times (4 - 1 - 1)$ and hence the acquaintance pruning strategy stops selecting a vertex from $V_A$. Therefore, we return the optimal solution $\{v_2, v_3, v_4, v_7\}$ with the total social distance as 62. □

---
[3]Some small modifications are made for better illustration.
[4]To find $A(V_S \cup \{v_2\})$, we derive $|V_A \cap N_{v_7}| + (k - |V_S - \{v_7\} - N_{v_7}|) = 4 + (1 - 0) = 5$ and $|V_A \cap N_{v_2}| + (k - |V_S - \{v_2\} - N_{v_2}|) = 2 + (1 - 0) = 3$, and then choose the smaller one. Therefore, we have $A(V_S \cup \{v_2\}) = 3$.
[5]To calculate $U(V_S \cup \{v_2\})$, we first need to consider the value of $|V_S - \{v_7\} - N_{v_7}|$ and $|V_S - \{v_2\} - N_{v_2}|$, and then choose the larger one. Since $|V_S - \{v_7\} - N_{v_7}| = |\emptyset| = 0$ and $|V_S - \{v_2\} - N_{v_2}| = |\emptyset| = 0$, we have $U(V_S \cup \{v_2\}) = 0$.

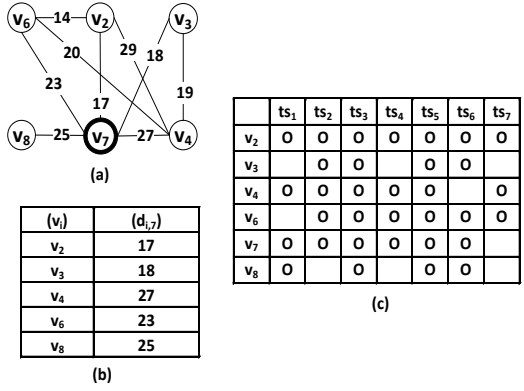

**Figure 3: Another illustrative example for SGQ and STGQ. (a) The sample social network, (b) the social distances of candidate attendees and (c) the schedules of candidate attendees.**

EXAMPLE 3. In this illustrating example for STGSelect, we extend the SGQ in Example 2 by considering the length of activity time as 3 (i.e., $m = 3$). When processing an STGQ, the schedules of candidate attendees provided in Figure 3(c) should be considered as well. Since $m = 3$, $ts_3$ and $ts_6$ are selected to be pivot time slots. For the first pivot time slot $ts_3$, $V_S = \{v_7\}$ and $V_A = \{v_2, v_3, v_4, v_6, v_8\}$ in the beginning. As obtained in Example 2, both of the exterior expansibility condition and the interior unfamiliarity condition hold when selecting $v_2$. Note that STGSelect also evaluates the temporal extensibility condition when selecting a vertex to ensure the feasibility in the temporal dimension. Since $(m-1)\left[\frac{p-|V_S \cup \{v_2\}|}{p}\right]^\phi = 2 \times (\frac{2}{4})^2 = \frac{1}{2}$ (assume $\phi = 2$) and $X(V_S \cup \{v_2\}) = 2$[7], the temporal extensibility condition also holds, and hence we can select $v_2$ from $V_A$ to $V_S$. Now we have $V_S = \{v_2, v_7\}$ and $V_A = \{v_3, v_4, v_6, v_8\}$. The later vertex selection ordering is identical to Example 2 since there is no violation on the temporal constraint, and we also obtain the first feasible solution $\{v_2, v_4, v_6, v_7\}$ (total social distance = 64) available in the activity period $[ts_2, ts_4]$. Until we select $v_3$ in the state $V_S = \{v_2, v_4, v_7\}$ and $V_A = \{(v_3)\}$, we find out that the temporal extensibility condition does not hold when selecting $v_3$, and then we increase $\phi$ since there is no other vertex in $V_A$ that we can choose. However, since $X(V_S \cup \{v_3\}) = 2 - 3 = -1$, the temporal extensibility condition does not hold even when the RHS of the inequality approaches 0. Therefore, we can remove $v_3$ and backtrack to the state $V_S = \{v_7\}$ and $V_A = \{v_3, v_4, v_6, v_8\}$. As shown in Example 2, the later branches violate the social constraints and hence lead to no feasible group. Therefore, $\{v_2, v_4, v_6, v_7\}$ is the only feasible group available in activity periods extended from the pivot time slot $ts_3$.

---

[6]There is no parenthesis for $v_3$, because we leave the branch of $V_S = \{v_2, v_7\}$.

Next, we start processing the second pivot time slot, i.e., $ts_6$. Different from $ts_3$, we have $V_S = \{v_7\}$ and $V_A = \{v_2, v_3, v_6, v_8\}$ in the beginning. Since $v_4$ is not available in the pivot time slot, we can directly remove it without further consideration. We then obtain $V_S = \{v_2, v_7\}$ and $V_A = \{v_3, v_6, v_8\}$ because selecting $v_2$ violates no constraint. Note that the LHS of the availability pruning condition is $\bar{t}_A^+(|V_A| - p + |V_S| + 1) - \bar{t}_A^-(|V_A| - p + |V_S| + 1) = \bar{t}_A^+(3 - 4 + 2 + 1) - \bar{t}_A^-(3 - 4 + 2 + 1) = \bar{t}_A^+(2) - \bar{t}_A^-(2)$. Since there are 2 vertices, i.e., $v_3$ and $v_8$, in $V_A$ not available in $ts_4$, $\bar{t}_A^-(2) = 4$. Besides, there are 2 vertices, i.e., $v_3$ and $v_8$, in $V_A$ not available in $ts_7$, $\bar{t}_A^+(2) = 7$. Therefore, the availability pruning condition holds since $\bar{t}_A^+(2) - \bar{t}_A^-(2) = 7 - 4 \leq m$, and we can stop selecting vertices from $V_A$ to $V_S$. Then we backtrack one step to the state $V_S = \{v_7\}$ and $V_A = \{v_3, v_6, v_8\}$. We can skip this final branch since the acquaintance pruning condition holds. Therefore, there exists no feasible group available in activity periods extended from $ts_6$. Finally, we return the group $\{v_2, v_4, v_6, v_7\}$ and the time period $[ts_2, ts_4]$ as the optimal result. □

## B. PROOFS

### B.1 Proof of Theorem 1

In the following, we prove that SGQ is NP-hard with the reduction from problem $\bar{k}$-plex, which is NP-hard [11]. A $\bar{k}$-plex with $c$ vertices is a subgraph in a graph $\overline{G} = (\overline{V}, \overline{E})$, such that every vertex in the subgraph can share no edge with at most $\bar{k} - 1$ other vertices in the subgraph. Problem $\bar{k}$-plex with parameters $c$ and $\bar{k}$ is to decide if $\overline{G}$ has a $\bar{k}$-plex with $c$ vertices. We prove that SGQ is NP-hard with the reduction from problem $\bar{k}$-plex. Specifically, we construct a weighted graph $G(V, E)$ by letting $V$ as $\{q\} \cup \overline{V}$ and letting $E$ as $\overline{E}$. We then add $c$ edges into $E$ to make $q$ connect to all the other vertices. The distance of every edge is 1. In the following, we prove that there exists a feasible group $F$ with $c + 1$ attendees in $G$ for SGQ with $s = 1$ and $k = \bar{k} - 1$, if and only if there exists a $\bar{k}$-plex with $c$ vertices in $\overline{G}$.

We first prove the sufficient condition. If we remove $q$ from $F$, every vertex in $F$ still shares no edge with at most $\bar{k} - 1$ other vertices, since $q$ connects to all the other vertices. In other words, $F - \{q\}$ with $c$ vertices corresponds to a $\overline{k}$-plex with size $c$ in $\overline{G}$. We then prove the necessary condition. If there exists a $\overline{k}$-plex with size $c$ in $\overline{G}$, $F$ with $q$ and the $c$ vertices must all satisfy the social radius constraint with $s$ as 1, since all the $c$ vertices connect to $q$ in $G$. Moreover, since each vertex in the $\overline{k}$-plex shares no edge with at most $\overline{k} - 1$ other vertices, $F$ with $q$ and the $c$ vertices must satisfy the acquaintance constraint. The theorem follows.

---

[7]$T_s = \{ts_1, ts_2, ts_3, ts_4, ts_5\}$ since $v_2$ and $v_7$ are available in them. Hence $|T_s| = 5$ and $X(V_S \cup \{v_2\}) = 5 - 3 = 2$.

## B.2 Proof of Theorem 2

In radius graph extraction, each of the removed vertices has no path with at most $s$ edges connecting to $q$, and no feasible solution thereby contains these vertices. Algorithm SGSelect includes three strategies: access ordering, distance pruning, and acquaintance pruning. For each $V_S$, the interior unfamiliarity condition only decides the order of selecting a vertex from $V_A$, while Lemma 1 shows that the exterior expansibility does not consider the vertex violating the acquaintance constraint. After we choose a vertex $u$ from $V_A$, Lemma 2 shows that the distance pruning specifies a lower bound on the total social distance that is derived from $V_A$. Therefore, the distance pruning will prune off only the solution with a larger total social distance. Moreover, Lemma 3 shows that the acquaintance pruning specifies the lower bound on the total inner degree on any set of vertices extracted from $V_A$ in any feasible solution. If we choose the required number of vertices with the largest inner degrees from $V_A$ and the result cannot exceed the above lower bound, the connectivity is too small for $V_A$ to obtain a feasible solution. The theorem follows.

## B.3 Proof of Theorem 3

Each pivot time slot is separated from a neighbor pivot time slot with $m-1$ time slots. Therefore, Lemma 4 shows that any feasible solution must include exactly one pivot time slot. In addition, the proposed algorithm considers the interval with $2m-1$ slots for each pivot time slot, and we derive the best solution by extending Algorithm SGSelect with the temporal extensibility. Moreover, Lemma 5 shows that the availability pruning discards $V_A$ only when there exists no feasible solution satisfying the availability constraint by incorporating any vertex from $V_A$. The solution obtained by Algorithm STGSelect is optimal because the algorithm chooses the pivot time slot and the corresponding group with the smallest total social distance at the end of the algorithm. The theorem follows.

## C. PSEUDO CODES

**Algorithm 1** SGSelect

**Require:** Graph $G(V, E)$, activity size $p$, social radius constraint with size $s$, and acquaintance constraint with size $k$
**Ensure:** Optimal group $F$
1: $d_{q,q}^0 = 0$, $d_{u,q}^0 = \infty$ for $u \neq q$;
2: **for** $i = 1$ to $s$ **do**
3:    $d_{q,q}^i = 0$;
4:    **for all** vertex $u \neq q$ in $V$ **do**
5:       $d_{u,q}^i = d_{u,q}^{i-1}$;
6:       **for all** vertex $v$ in $N_u$ **do**
7:          **if** $d_{v,q}^{i-1} + c_{u,v} < d_{u,q}^i$ **then**
8:             $d_{u,q}^i = d_{v,q}^{i-1} + c_{u,v}$;
9: Extract all vertices $w$ in $V$ with $d_{w,q}^s < \infty$ and form the set $V_F$;
10: $V_S = \{q\}, V_A = V_F - \{q\}, TD = \infty, D = \infty, F = \emptyset$;
11: $ExpandSG(V_S, V_A, TD)$;
12: **if** $D \neq \infty$ **then**
13:    Output $F$;
14: **else**
15:    Output "Failure";

**Algorithm 2** ExpandSG

**Function:**
$ExpandSG(inV_S, inV_A, inTD)$
1: $V_S = inV_S, V_A = inV_A, TD = inTD$;
2: **while** $|V_S| + |V_A| \geq p$ **do**
3:    **if** there is any unvisited vertex in $V_A$ **then**
4:       Select an unvisited vertex $u$ with the minimum social distance to $q$ and mark $u$ as visited;
5:    **else if** $\theta > 0$ **then**
6:       Reduce $\theta$ and mark remaining vertices in $V_A$ as unvisited;
7:    **else**
8:       BREAK;
9:    **if** $u$ satisfies the exterior expansibility condition **then**
10:      **if** $u$ satisfies the interior unfamiliarity condition **then**
11:         $V_S = V_S + \{u\}, V_A = V_A - \{u\}, TD = TD + d_{u,q}$;
12:         **if** the distance pruning condition holds for $TD$ **or** the acquaintance pruning condition holds for $V_A$ **then**
13:            BREAK;
14:         **else if** $|V_S| < p$ **then**
15:            $ExpandSG(V_S, V_A, TD)$;
16:         **else**
17:            $D = TD, F = V_S$;
18:            BREAK;
19:      **else if** $\theta = 0$ **then**
20:         $V_A = V_A - \{u\}$;
21:    **else**
22:       $V_A = V_A - \{u\}$;

**Algorithm 3** STGSelect

**Require:** Graph $G(V, E)$, activity size $p$, social radius constraint with size $s$, acquaintance constraint with size $k$, and activity length $m$
**Ensure:** Optimal group $F$ and activity period $P$
1: $d_{q,q}^0 = 0$, $d_{u,q}^0 = \infty$ for $u \neq q$;
2: **for** $i = 1$ to $s$ **do**
3:    $d_{q,q}^i = 0$;
4:    **for all** vertex $u \neq q$ in $V$ **do**
5:       $d_{u,q}^i = d_{u,q}^{i-1}$;
6:       **for all** vertex $v$ in $N_u$ **do**
7:          **if** $d_{v,q}^{i-1} + c_{u,v} < d_{u,q}^i$ **then**
8:             $d_{u,q}^i = d_{v,q}^{i-1} + c_{u,v}$;
9: Extract all vertices $w$ in $V$ with $d_{w,q}^s < \infty$ and form the set $V_F$;
10: Select a pivot time slot every $m$ time slots, i.e., select time slots with ID as $im$, where $i$ is a positive integer;
11: **while** there is any unprocessed pivot time slot $im$ **and** $q$ is available in $im$ **do**
12:    $V_S = \{q\}, V_A = V_F - \{q\} - \{$vertices not available in $im\}, TD = \infty, D = \infty, F = \emptyset, TP = [(i-1)m+1, (i+1)m-1], P = \emptyset$;
13:    $ExpandSTG(V_S, V_A, TD, TP, im)$;
14:    Remove pivot time slot $im$;
15: **if** $D \neq \infty$ **then**
16:    Output $F$ and $P$;
17: **else**
18:    Output "Failure";

## Algorithm 4 ExpandSTG

**Function:**
$ExpandSTG(inV_S, inV_A, inTD, inTP, inIM)$

1: $V_S = inV_S$, $V_A = inV_A$, $TD = inTD$, $TP = inTP$, $im = imIM$;
2: **while** $|V_S| + |V_A| \geq p$ **do**
3:   **if** there is any unvisited vertex in $V_A$ **then**
4:     Select an unvisited vertex $u$ with the minimum social distance to $q$ and mark $u$ as visited;
5:   **else**
6:     **if** $\theta > 0$ **then**
7:       Reduce $\theta$ and mark remaining vertices in $V_A$ as unvisited;
8:     **else if** $\phi <$ a predetermined threshold $t$ **then**
9:       Increase $\phi$ and mark remaining vertices in $V_A$ as unvisited;
10:      **if** $\phi \geq t$ **then**
11:        set the RHS of the temporal extensibility condition as 0;
12:    **else**
13:      BREAK;
14:  **if** $u$ satisfies the exterior expansibility condition **then**
15:    **if** $u$ satisfies the interior unfamiliarity condition **then**
16:      **if** $u$ satisfies the temporal extensibility condition **then**
17:        $V_S = V_S + \{u\}$, $V_A = V_A - \{u\}$, $TD = TD + d_{u,q}$;
18:        Update $TP$ using the available time slots of $u$;
19:        **if** the distance pruning condition holds for $TD$ **or** the acquaintance pruning condition holds for $V_A$ **or** the availability pruning condition holds for $V_A$ **then**
20:          BREAK;
21:        **else if** $|V_S| < p$ **then**
22:          $ExpandSTG(V_S, V_A, TD, TP, im)$;
23:        **else**
24:          $D = TD$, $F = V_S$, $P = TP$;
25:          BREAK;
26:      **else if** $\phi \geq t$ **then**
27:        $V_A = V_A - \{u\}$;
28:    **else if** $\theta = 0$ **then**
29:      $V_A = V_A - \{u\}$;
30:  **else**
31:    $V_A = V_A - \{u\}$;

## D. INTEGER PROGRAMMING FORMULATION OF STGQ

In the following, we model STGQ as an Integer Programming problem, where the formulation can support SGQ by removing the constraints for the temporal dimension. The derived model, together with any Integer Programming commercial solver, such as CPLEX [4], can obtain the optimal solution.

We first define a number of decision variables in the formulation. Let $\delta_u$ denote the social distance from $q$ to $u$, $\delta_u \geq 0$. Let binary variable $\phi_u$ denote whether vertex $u$ is in $F$. Let binary variable $\pi_{u,i,j}$ denote if $e_{i,j}$ is located in the shortest-path from $q$ to $u$, where $i$ is ascendant to $j$ in the path. Let binary variable $\tau_t$ denote if the activity starts in slot $t$. The problem is to minimize the total social distance from $q$ to all vertices in $F$, i.e.,

$$\min \sum_{u \in F} \delta_u.$$

However, because $F$ needs to be decided in the problem, too, the above function becomes non-linear thus cannot act as the objective function for STGQ. To address the above problem, we propose an alternative linear objective function, i.e.,

$$\min \sum_{u \in V} \delta_u,$$

which considers set $V$ instead of $F$. Therefore, to obtain the correct objective value, we need to design a constraint that ensures that $\delta_u = 0$ for every $u$ not in $F$. Specifically, the proposed formulation includes the following constraints.

$$\sum_{u \in V} \phi_u = p. \quad (1)$$
$$\phi_q = 1. \quad (2)$$
$$\sum_{v \in N_u} \phi_v \geq (p-1)\phi_u - k, \forall u \in V. \quad (3)$$
$$\sum_{i \in N_q} \pi_{u,q,i} = \phi_u, \forall u \in V, u \neq q. \quad (4)$$
$$\sum_{i \in N_u} \pi_{u,i,u} = \phi_u, \forall u \in V, u \neq q. \quad (5)$$
$$\sum_{i \in N_j} \pi_{u,i,j} - \sum_{i \in N_j} \pi_{u,j,i} = 0,$$
$$\forall u, j \in V, u \neq q, j \neq q, \text{ and } j \neq u. \quad (6)$$
$$\sum_{e_{i,j} \in E} c_{i,j} \pi_{u,i,j} = \delta_u, \forall u \in V. \quad (7)$$
$$\sum_{e_{i,j} \in E} \pi_{u,i,j} \leq s, \forall u \in V, u \neq q. \quad (8)$$
$$\sum_{1 \leq t \leq T-m+1} \tau_t = 1, \forall u \in V, u \neq q. \quad (9)$$
$$\phi_u \leq 1 - \tau_t + a_{u,\hat{t}},$$
$$\forall u \in V, 1 \leq t \leq T-m+1, \text{ and } t \leq \hat{t} \leq t+m-1. \quad (10)$$

Constraint (1) guarantees that exactly $p$ vertices are selected in solution set $F$, and vertex $q$ must be one of them as stated in constraint (2), to ensure that an activity with size $p$, including the initiator, can be obtained. Constraint (3) specifies the acquaintance constraint and guarantees that every vertex $u$ in $F$, i.e., $\phi_u = 1$, can have no social link with at most $k$ vertices. In other words, at least $p - 1 - k$ neighbors of $u$ must belong to $F$, where 1 corresponds to $u$ itself. If $u$ is not in $F$, $\phi_u = 0$, we let constraint (3) become non-enforcing because this constraint does not restrict the value of $\phi_v$ for each neighbor $v$ of $u$.

For each vertex $u$, constraints $(4) - (8)$ find a path from $q$ to $u$ and the corresponding distance, if $u$ is chosen in $F$, i.e., $\phi_u = 1$. Specifically, if $\phi_u$ is 1, constraints (4) and (5) select an incident edge for $q$ and $u$ to be included in the path, respectively. In this case, for every other vertex $j$, constraint (6) ensures that either $j$ is not in the path, or $j$ has one ascendant and one descendant neighbor vertices in the path. Constraint (7) then derives $\delta_u$ according to the total distance in the path, i.e., $\sum_{e_{i,j} \in E} c_{i,j} \pi_{u,i,j}$, and the objective function of the formulation enforces that the path must be the shortest one. If $\phi_u$ is 0, constraints (4) and (5) do not create the path for $u$, and the distance $\delta_u$ appearing in the objective function is thereby 0 according to constraint (7). Constraint (8) states the social radius constraint and ensures that the path cannot include more than $s$ edges.

The last two constraints find the time slot in which the activity begins, and the proposed formulation can support SGQ by discarding the two constraints. Constraint (9) specifies that the activity must begin in no later than slot $T-m+1$, where $T$ is the largest ID of a time slot; otherwise, the activity must span fewer than $m$ slots. The last constraint prohibits a candidate attendee $u$ from joining the activity if $u$ is not available in the corresponding period. Specifically, if the activity starts at $t$ and spans the slots from $t$ to $t+m-1$, i.e., $\tau_t = 1$, $\phi_u$ must be 0 if $u$ is not available, i.e., $a_{u,\hat{t}}$ is 0, for any slot $\hat{t}$ in the period. If $a_{u,\hat{t}} = 1$, $u$ is allowed, but not enforced, to join the activity. In contrast, if the activity does not start from $t$, i.e., $\tau_t = 0$, we let this constraint become a non-enforcing one because $\phi_u$ is not restricted in this case.